\documentclass[letterpaper, 10 pt, conference]{ieeeconf}  

\IEEEoverridecommandlockouts                        
\usepackage{amsthm}
\usepackage{amsmath}
\usepackage{amsfonts}
\usepackage{graphicx}
\usepackage{algorithm}
\usepackage{caption}
\usepackage{subcaption}
\usepackage[noend]{algpseudocode}
\usepackage{cite}
\usepackage{balance}
\usepackage{bm}
\usepackage{xcolor}
\usepackage{float}
\usepackage{nicefrac}
\usepackage{comment}
\usepackage{booktabs}
\usepackage{colortbl}
\usepackage{xcolor, soul}

\newtheorem{problem}{Problem}

\newcommand{\real}{\mathbb{R}}

\DeclareMathOperator*{\argmin}{arg\,min}

\definecolor{DarkGray}{rgb}{0.19,0.31,0.31}

\title{\LARGE \bf
A Fisher Information based Receding Horizon Control Method for Signal Strength Model Estimation
}

\author{Yancheng Zhu$^1$ and Sean B. Andersson$^{1,2}$
\\
$^1$Dept. of Mechanical Engineering, $^2$Division of Systems Engineering,\\ 
Boston University, Boston, MA 02215, USA \\%
}

\begin{document}

\maketitle
\thispagestyle{empty}
\pagestyle{empty}

\begin{abstract}

This paper considers the problem of localizing a set of nodes in a wireless sensor network when both their positions and the parameters of the communication model are unknown. We assume that a single agent moves through the environment, taking measurements of the Received Signal Strength (RSS), and seek a controller that optimizes a performance metric based on the Fisher Information Matrix (FIM). We develop a receding horizon (RH) approach that alternates between estimating the parameter values (using a maximum likelihood estimator) and determining where to move so as to maximally inform the estimation problem. The receding horizon controller solves a multi-stage look ahead problem to determine the next control to be applied, executes the move, collects the next measurement, and then re-estimates the parameters before repeating the sequence. We consider both a Dynamic Programming (DP) approach to solving the optimal control problem at each step, and a simplified heuristic based on a pruning algorithm that significantly reduces the computational complexity. We also consider a modified cost function that seeks to balance the information acquired about each of the parameters to ensure the controller does not focus on a single value in its optimization. These approaches are compared against two baselines, one based on a purely random trajectory and one on a greedy control solution. The simulations indicate our RH schemes outperform the baselines, while the pruning algorithm produces significant reductions in computation time with little effect on overall performance.
\end{abstract}

\section{Introduction}
\label{sec:intro}

Wireless Sensor Networks (WSNs) are composed of a collection of (possibly mobile) sensors deployed over a geographical area and that collect data about their surroundings over time. Such systems have applications in a broad range of settings, including environmental monitoring~\cite{HART:Reviews}, smart cities~\cite{Lei:WSNSmartCity}, disaster and rescue missions~\cite{Bhosle:WSNForestdisaster}, and surveillance~\cite{Zhijun:Motionplanningsurveillance}. To be useful, however, their data needs to be extracted and brought back to a data sink to be accessed by users interested in the information acquired by the WSN. When the WSN is deployed over a large and remote area, the use of mobile agents to act as data mules can be an effective solution to this data harvesting task~\cite{zhu:2022optimalcontroldataharvesting}. In order to optimize the data collection process, the agents need to know both where the sensors are located and the parameters that define the communication. While GPS can be used to localize, it is not always available and when power resources on the sensor nodes are limited, it is important to minimze the amount of information that needs to be communicated from node to agent. In this work we utilize the Received Signal Strength (RSS) since this signal is automatically present if there is communication taking place.

The use of the RSS for localization of transmitting nodes brings multiple benefits, including a relatively low deployment cost and compatibility with WiFi or Bluetooth Low Energy (BLE) enabled devices~\cite{plets2019joint}. Typically, methods that use RSS apply a Line-of-Sight (LoS) path-loss model to describe signal propagation in outdoor environments~\cite{al2014modeling}. Since these parameters depend on the specifics of a particular environment and may vary over time, the problem of network localization based on RSS measurements obtained by agents where the parameters of the signal propagation model are unknown is also of interest. Approaches include a map-based iterative learning control method~\cite{esrafilian2020three}, cooperative localization using a dynamic model~\cite{liang2015received}, and distance-based localization~\cite{suroso2020distance}. However, due to the presence of disturbances such as multi-path fading in signal propagation~\cite{zimmermann1999multi} as well as measurement noise, the RSS signal can fluctuate. Inspired by this, our goal in this work is to devise a method that enables the active planning of the agent's motion to continually update the estimation and to plan a trajectory that maximally informs estimation of the parameters.

There are a variety of ways to represent the information acquired in a collection of measurements. Because we are using a parametric model, we choose to use Fisher Information (FI), seeking to maximize a measure derived from the Fisher Information Matrix (FIM) to drive the optimization. One of the benefits of using the FIM is that it is independent of any particular estimation scheme and maximizing the FIM is equivalent to minimizing the Cram\'{e}r-Rao Lower Bound (CRLB), a fundamental limit on the variance of any unbiased estimator. The FIM has been used in a similar way in active planning for aerial photogrammetry~\cite{lim2023fisher}, in vision-based small UAV navigation~\cite{frew2006adaptive}, and in real-time single particle tracking control~\cite{vickers2023information}. In these control problems, FIM serves as a tool to quantify the information acquired.

The optimal control problem we define is related to work in trajectory design for wireless localization systems~\cite{esrafilian2020three,liang2015received}. We start with a RSS model that relies on a LoS path-loss model with power measurements that are corrupted by noise that follows a Gaussian distribution. We consider a single agent tasked with estimating the unknown parameters in the signal propagation model and the locations of multiple sensors in a three dimensional outdoor environment. Because the FIM is conditioned on knowledge of the parameters, we use a receding horizon (RH) control approach, alternating between estimating all parameters and locations using Maximum Likelihood Estimation (MLE) based on all measurements acquired so far, and optimizing the locations for the next measurements based on the FIM.

The contributions of our work consist of: (1) the formulation of an optimization problem to minimize the estimation error  for RSS model in a WSN by applying an active planning method; (2) the development of a FIM-based RH controller for agent motion planning in real time; (3) a detailed analysis of the Fisher information distribution in the space for unknown parameters in the signal strength model; (4) a modified cost function that seeks to balance the information across all parameters while still using a fairly simple cost function derived from the FIM.

\section{Problem Formulation}
\label{sec:formulation}
We begin this section by introducing the system dynamics and signal strength path-loss model, and then formally define the online estimation and control problem. 

\begin{figure}[htp]
	\centering
	\includegraphics[width=0.8\columnwidth]{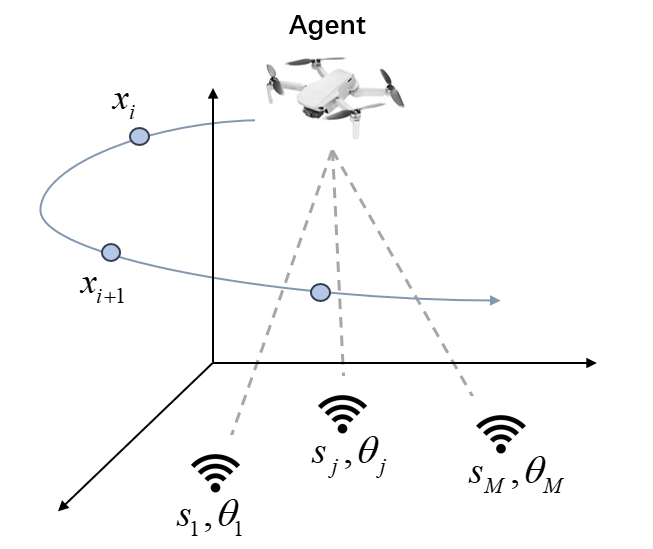}
	\caption{Sketch of the problem setup. A single agent flies over an environment while communicating with a collection of sensor nodes distributed on the ground.}
	\label{fig:signal_measuring_schematic}
\end{figure}

We consider the scenario illustrated in Fig.~\ref{fig:signal_measuring_schematic}. There is a collection of $M$ sensor nodes, located at positions $s_j \in \real^3$, $j=1,\dots,M$, on non-level ground in an outdoor environment. Each node communicates with a single agent flying above. The position of the agent at time step $i$ is given by $x_i \in \real^3$ and evolves according to
\begin{align}
    x_{i+1} = x_i + u_i, \quad u_i \in \mathcal{U},
    \label{eq:agentDynamics}
\end{align}
where $u_i$ is the control input and $\mathcal{U}$ is a set of admissible controls. We describe the measured RSS from sensor node $j$ by the agent at $x_i$ in dB, using the LoS signal pass loss model in \cite{al2014modeling} as
\begin{equation} 
    \label{eq:signal_path_model}
    y_{ij}(x_i,s_j) =  K_j -\gamma_j\log_{10}(d(x_i,s_j)) + \xi_j,
\end{equation}
where $\xi_j$ is the measurement noise and $d(x_i,s_j)$ is the Euclidean distance.

The parameters $\gamma_j$ and $K_j$ in the signal path-loss model, as well as the locations of the nodes, are unknown and need to be estimated. We collect those parameters into a single vector $\bm{\theta} \in \real^{5M}$ as 
\begin{equation}
    \bm{\theta}=\begin{bmatrix}\theta_1^T & \dots &\theta_M^T\end{bmatrix}^T, \quad \theta_j = \begin{bmatrix} \gamma_j & K_j & s_j^T\end{bmatrix}^T.
\end{equation}

The goal is to collect measurements to optimally estimate $\bm{\theta}$ from the measured RSS. ``Optimal'' here is defined as the set of measurements that yield the best estimate of the parameters. We formalize this notion using the posterior CRLB~\cite{bergman2001posterior-carmer-Rao}, which states that the Mean Squared Error (MSE) of the estimated parameters is lower bounded by a function of the FIM, namely
\begin{equation} 
    \label{eq: MSE for FI} 
    \mathrm{MSE}(\bm{\hat{\theta}}) \geq \mathrm{tr}(\rm{F}_{N}^{-1}(\bm{\theta})),
\end{equation}
where $\hat{\bm{\theta}}$ is (any) unbiased estimate and $F_N$ is the FIM after $N$ measurements. This leads to the following formal problem statement. 
\begin{problem}
\label{prob:basic_online_control_FIM}
\begin{equation*}
\begin{array}{cl}
    \label{eq:optimization_problem_formulation}
    \min_{u_i} & J_1=  \rm{tr}(\rm{F}^{-1}_{N}(\bm{\hat{\theta}})) \\
    \\ {\textrm{ subj. to }} & x_{i+1} = x_i + u_i, 
    \\ {} & y_i = \Phi ( x_i, \bm{\theta}),
    \\ {} & \bm{\hat{\theta}}_{i+1} = \Theta( x_i, y_i,  \bm{\hat{\theta}}_i),
    \\ {} & \rm{F}_{N}(\bm{\hat{\theta}}) = \sum_{i=1}^{N}\rm{F_{i}}(x_i,\bm{\hat{\theta_i}}), 
    \\ {} & i=1,\dots,N.
\end{array}
\end{equation*}
\end{problem}

Here $y_i \in \real^M$ is the collection of RSS measurements, $\Phi(x_i, \bm{\theta})$ is the combined measurement function with each element given by \eqref{eq:signal_path_model}, $\Theta(x_i,y_i,\hat{\bm{\theta}}_i)$ is a given estimator, and $F_i(x_i,\hat{\bm{\theta}}_i)$ is the FIM from measurement $i$.

\section{Method}
\label{sec:parameter_opt_control}

\begin{figure}[htp]
	\centering
	\includegraphics[width=0.42\textwidth]{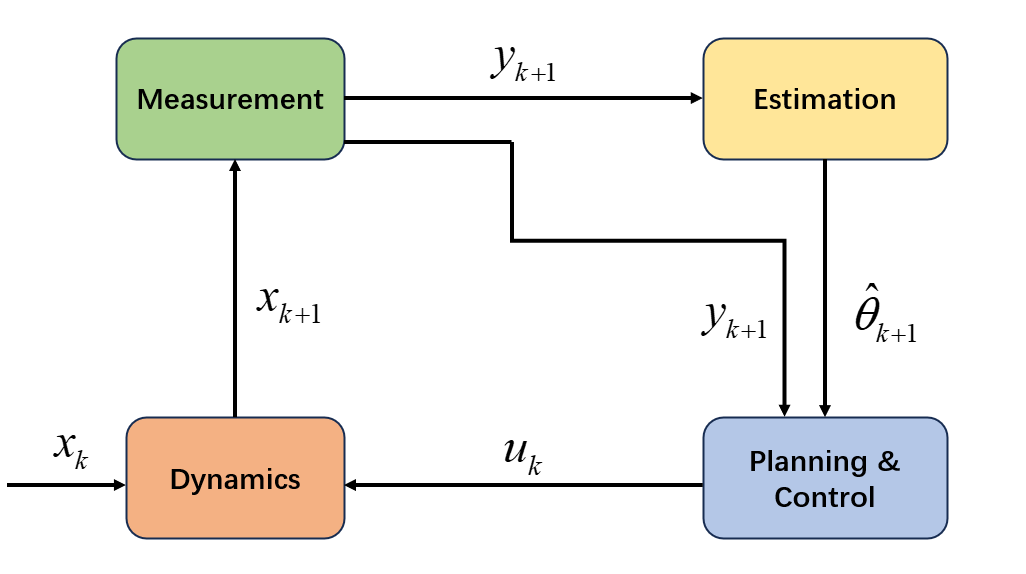}
	\caption{Block diagram of the estimation and control framework for active motion planning.}
	\label{fig:online_control_estimation_framework}
\end{figure}

Our general approach to Problem \ref{prob:basic_online_control_FIM}, illustrated in Fig.~\ref{fig:online_control_estimation_framework}, is that of receding horizon control where we alternate between updating the estimate of the parameters based on all the data collected so far, and planning where to take the remaining measurements based on that most recent estimate.

In this section, we describe the details in each of the stems, beginning with the ML estimator.

\begin{figure*}[htbp!]
    \centering
    \begin{subfigure}{0.245\textwidth}
        \centering\includegraphics[width=\textwidth]{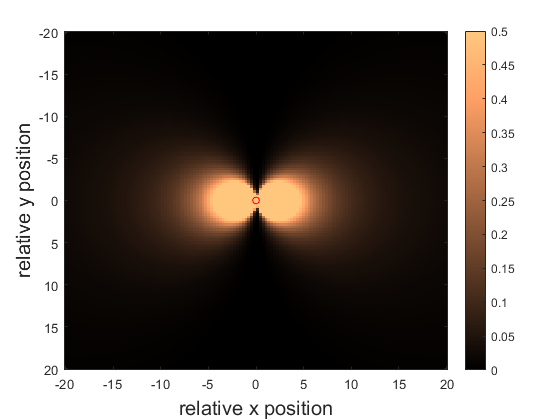}
        \caption{FIM for $s^x_j$}
        \label{fig:FIM_X}
    \end{subfigure}
    \begin{subfigure}{0.245\textwidth}
        \centering\includegraphics[width=\textwidth]{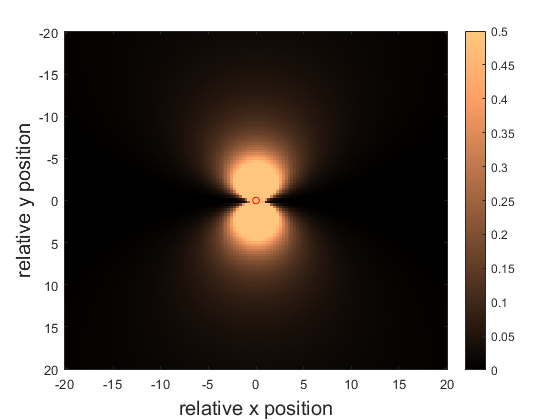}
        \caption{FIM for $s^y_j$}
        \label{fig:FIM_Y}
    \end{subfigure}
    \begin{subfigure}{0.245\textwidth}
        \centering\includegraphics[width=\textwidth]{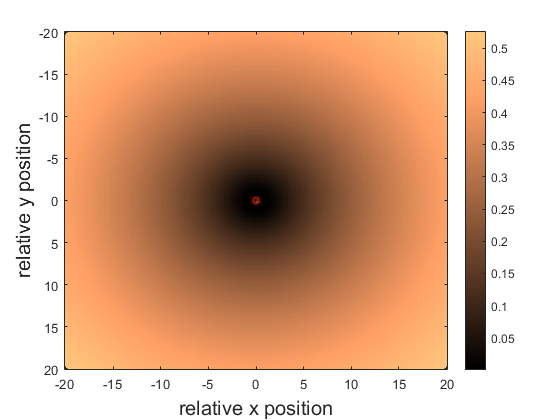}
        \caption{FIM for $\gamma_j$}
        \label{fig:FIM_Gamma}
    \end{subfigure}
    \begin{subfigure}{0.245\textwidth}
        \centering\includegraphics[width=\textwidth]{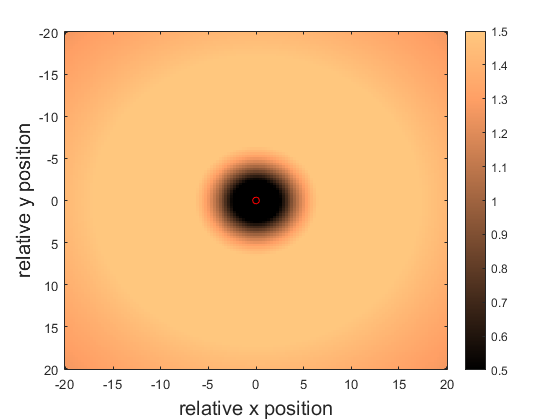}
        \caption{$\rm{tr}(\rm{F}^{-1}_{N}(\bm{\theta}))$}
        \label{fig:MSE_Lower_Bound}
    \end{subfigure}
    \caption{Illustration of FIM for the relative positions and parameter $\gamma$ for the $i$th sensor node. The light color in (a), (b) and (c) represents the high value of FIM, while the light area in (d) means the high value of MSE lower bound.}
    \label{fig:FIM_for_ith_sensor}
\end{figure*}

\subsection{ML Estimator \label{sec:MLEstimator }}

While there are many estimators that can be used, we choose to use MLE as it is both consistent (achieving unbiasedness in the limit of a large number of measurements) and efficient (achieving the CRLB in that limit)~\cite{mle_myung2003tutorial}. Given the Gaussian noise model on the measurement noise $\xi_i$, from \eqref{eq:signal_path_model} we have
\begin{equation} 
    \label{eq:distribution of y}
    p(y_{ij} | \mu_{ij},\sigma_j^2) = \frac{1}{\sqrt{2\pi}\sigma_j} \exp{ \left(- \frac{(y_{ij} - \mu_{ij})^2}{2\sigma_j^2} \right) },
\end{equation}
where $\mu_{ij}$ is defined as
\begin{equation} 
    \label{eq:presentation of mu}
    \mu_{ij} = K_j - \gamma_j\log_{10}(\|x_i-s_j\|_2).
\end{equation}
Taking the log of this yields
\begin{equation} 
    \label{eq:distribution of y in log} 
    \log p(y_{ij} | \mu_{ij},\sigma_j^2) = -\frac{1}{2}\log2\pi - \frac{1}{2}\log\sigma_j^2 - \frac{1}{2\sigma_j^2}(y_{ij}-\mu_{ij})^2.
\end{equation}

Under the assumption that the measurements are independent, then after $k$ measurements the joint log likelihood is given by
\begin{equation} 
    \label{eq:distribution of ml in log} 
    \log p(y_{11},\dots,y_{kM} | \bm{\theta}) = \sum_{i=1}^{k} \sum_{j=1}^{M} \log p(y_{ij} | \bm{\theta}).
\end{equation}
Inserting \eqref{eq:distribution of y in log} into \eqref{eq:distribution of ml in log} and eliminating all constant terms, the ML estimate of the parameters is given by
\begin{equation} 
    \label{eq: MLE and minimizing objective} 
    \hat{\bm{\theta}} = \argmin_{\bm{\theta}}\mathcal{L} =  \sum_{i=1}^{k} \sum_{j=1}^{M} \frac{1}{2\sigma_j^2}(y_{ij}-\mu_{ij})^2.
\end{equation}

We solve this nonlinear least squares problem using a conjugate gradient method~\cite{glowinski1985continuation}.

\subsection{Fisher Information Matrix \label{sec:Fisher_information}}

To calculate the FIM, we begin with the information from $y_{ij}$, the $i^{th}$ measurement of the RSS from node $j$, with the agent at position $x_i$. For simplicity of notation, we redefine \eqref{eq:distribution of y in log} as
\begin{equation}
    \label{eq:distribution of y in log recall} 
    f_{p_i}(x_i,\theta_j) = \log p(y_{ij} | \theta_j),
\end{equation}
where we have made explicit the dependence on the agent position. The FIM corresponding to this measurement is the expected value of the observed information, given as 
\begin{equation}   
    \label{eq:matrix Fij}
    \mathrm{F}_i(x_i,\theta_j)= \mathrm{E} \left[ \left( \frac{\partial{}}{\partial{\theta_j}}f_{p_i}(\theta_j)\right)   \left( \frac{\partial{}}{\partial{\theta_j}}f_{p_i}(\theta_j)\right) ^T \right],
\end{equation} 
where 
\begin{subequations}
    \label{eq:Ficomponents}
    \begin{align} 
        &\frac{\partial}{\partial K_j}(f_{p_i}) = \frac{1}{\sigma_j^2}(y_{ij}-\mu_{ij}), \label{eq:ddK} \\
        &\frac{\partial}{\partial \gamma_j}(f_{p_i}) = -\frac{1}{\sigma_j^2}(y_{ij}-\mu_{ij})\log_{10}\|x_i-s_j\|_2, \\
        &\frac{\partial}{\partial s^x_j}(f_{p_i}) = \frac{1}{\sigma_j^2}(y_{ij}-\mu_{ij}) \frac{\gamma_j(x_i^x-s^x_j)}{\|x_i-s_j\|^2_2\ln10}, \\
        &\frac{\partial}{\partial s^y_j}(f_{p_i}) = \frac{1}{\sigma_j^2}(y_{ij}-\mu_{ij}) \frac{\gamma_j(x_i^y-s^y_j)}{\|x_i-s_j\|^2_2\ln10}, \\
        &\frac{\partial}{\partial s^z_j}(f_{p_i}) = \frac{1}{\sigma_j^2}(y_{ij}-\mu_{ij}) \frac{\gamma_j(x_i^z-s^z_j)}{\|x_i-s_j\|^2_2\ln10}. 
    \end{align}
\end{subequations}
The combined FIM $\mathrm{F}_i(\bm{\hat{\theta}})$ for the measurements of the received power from all the sensor nodes is a diagonal matrix as
\begin{equation}   
    \label{eq:matrix Fi}
    \mathrm{F}_i(x_i,\bm{\theta})=  \text{diag} \left(\mathrm{F}_i(x_i,\theta_1) \dots  \mathrm{F}_i(x_i,\theta_M) \right).
\end{equation} 

As expressed in Problem \ref{prob:basic_online_control_FIM}, the goal is to maximize the information gain in terms of the FIM. Notice first from \eqref{eq:ddK} that the information gain relative to the gain $K_j$ does not depend on the relative position of the agent. The situation is more complicated for the other parameters. In Fig.~\ref{fig:FIM_for_ith_sensor} we show the values of the FIM with respect to each of the remaining parameters as a function of the relative position of the agent and sensor at a fixed agent height of 0. It is immediately clear that the best position to estimate each of the parameters is different; the best location for determining the location of the sensor in one axis yields no information about the other axis, and moving far away to best inform the fading parameter reduces the information about either position. Combining these into the single scalar cost that gives the lower bound on the overall MSE leads to the image in Fig.~\ref{fig:MSE_Lower_Bound}, showing that the best measurement in terms of the cost may not be particularly informative for any individual parameter. We will revisit this in Sec. \ref{sec:RH_controller}.

\begin{figure}[htp!]
    \begin{subfigure}{0.245\textwidth}
        \centering\includegraphics[width=\textwidth]{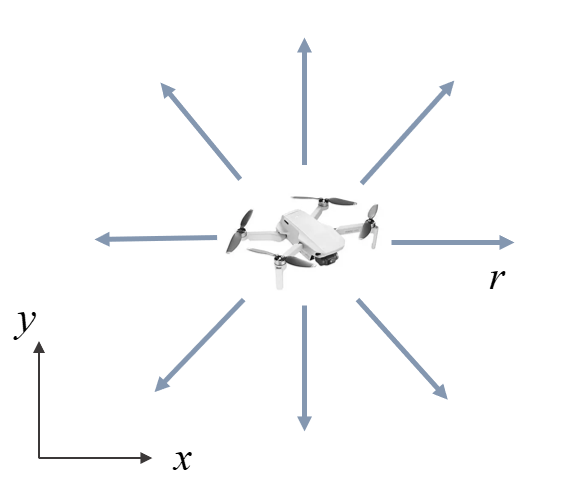}
        \caption{Horizontal control}
        \label{fig:horizontal_control}
    \end{subfigure}
    \begin{subfigure}{0.235\textwidth}
        \centering\includegraphics[width=\textwidth]{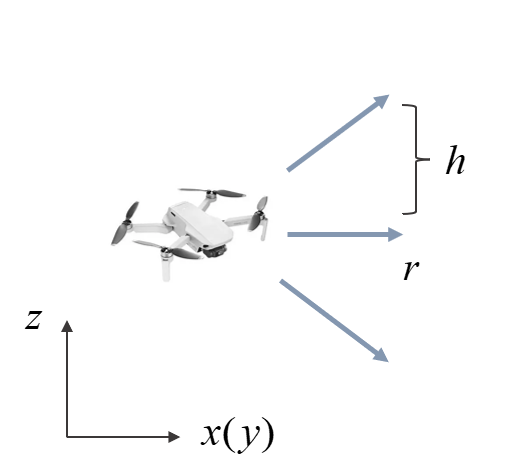}
        \caption{Vertical control}
        \label{fig:vertical_control}
    \end{subfigure}
    \caption{Demonstration of all possible control actions. (a) The vectors reflect eight horizontal control inputs. (b) The vectors reflect three vertical control inputs.}
    \label{fig:control_input_illustration}
\end{figure}

\subsection{Receding Horizon Controller Design\label{sec:RH_controller}}
To define our RH controller, we consider planning the motion of the agent over a future horizon of $T$ steps. To simplify the calculations, we limit ourselves to a finite number of control actions. As shown in Fig. \ref{fig:control_input_illustration}, we consider 24 possible moves, selected from eight possible horizontal moves with length $r$  and three possible vertical moves with height $h$. Note that it is straightforward to include a 0 control, but it is reasonable to the system that the agent will move to acquire new information and thus we omit that possibility. Let $k$ denote the current time step. The receding horizon control problem is then defined as follows.

\begin{problem}
\label{prob:RH_MPC_optimization}
\begin{equation*}
\begin{array}{cl}
    \label{eq:RH_optimization_problem_formulation}
    \min_{{u}_{k,1},\dots,{u}_{k,T}} & J_2 = \rm{tr}(\rm{F}^{-1}_{k+T}),\\
    \\ {\textrm{ subj. to }}  & \hat{x}_{k,i+1} = \hat{x}_{k,i} + u_{k,i}, 
    \\ {} & \hat{x}_{k,1} = x_k,
    \\ {} & u_{k,i} \in \mathcal{U},
    \\ {} & \rm{F}_{k+T} = \sum_{i=1}^T \lambda^{i}\cdot\rm{F}_{k,i}(\hat{x}_{k,i},\hat{\bm{\theta}}_k),
    \\ {} & i=1,\dots,T.
\end{array}
\end{equation*}
\end{problem}

Here $\mathcal{U}$ is the finite set of control values and $\lambda ( 0< \lambda \leq 1)$ is a user-defined discount factor which is motivated by the fact that the prediction of the value of the parameter becomes less certain in the future. Note that there exists a rare case of ill-conditioned $\rm{F}_{k+T}$, and we expanded our solution by incorporating a penalty to decrease the likelihood of this occurrence (see Sec.\ref{sec:Penalty}). 
This problem can be solved using Dynamic Programming (DP) for small values of $T$. As is usual in RH methods, we apply the first control and then repeat. While DP can be used, its computational complexity grows exponentially as ${O}(N_u^T)$ where $N_u$ is the number of control choices and thus this quickly becomes infeasible for even moderate $T$. We therefore introduce a pruning approach. 

\begin{figure}[htbp!]
    \centering\includegraphics[width=0.8\columnwidth]{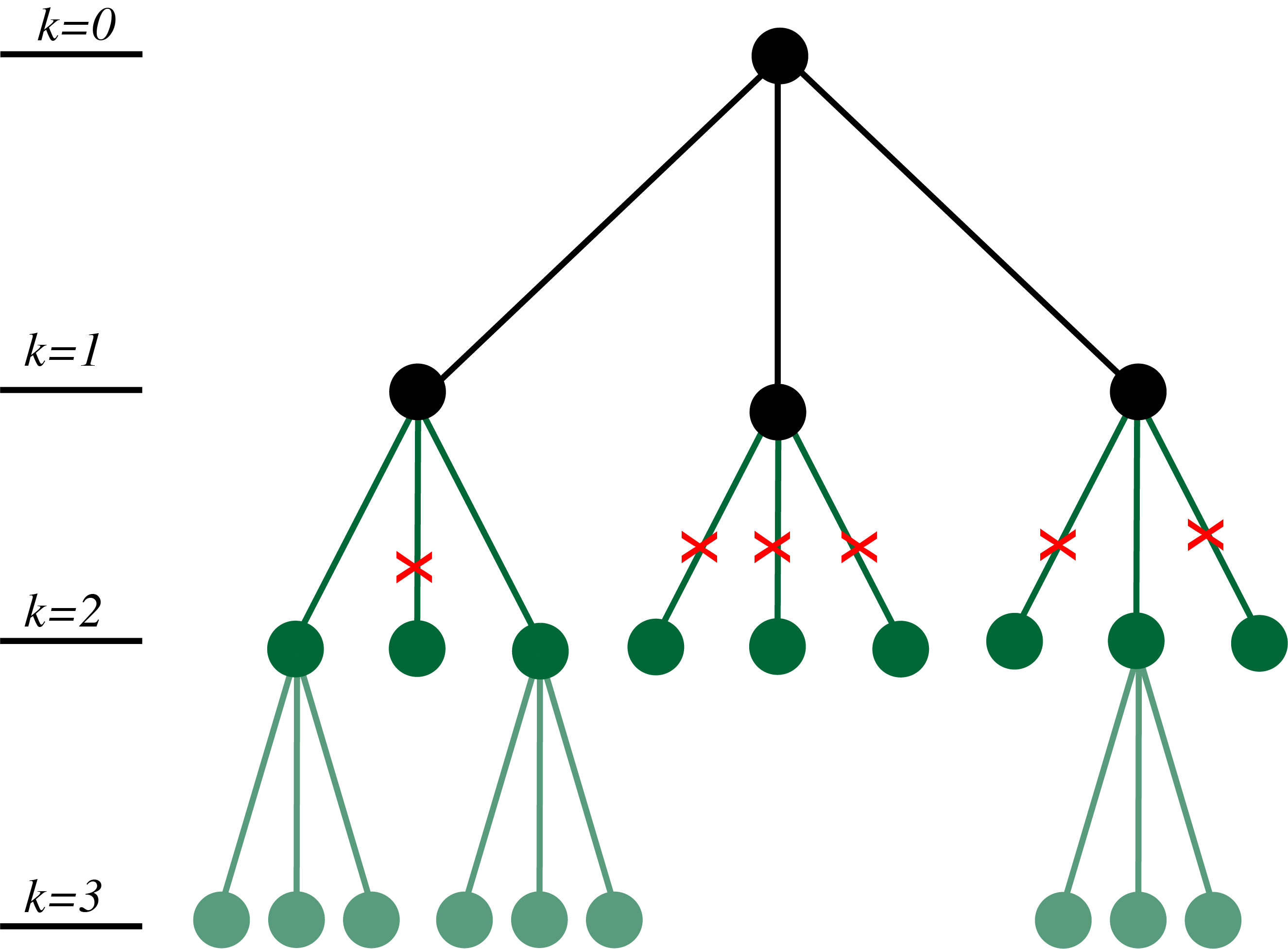}
    \caption{Sketch of the pruning algorithm with $N_u=3$ and $M_u=3$. At each stage there are only $N_uM_u$ actions to consider.}
    \label{fig:pruning}
\end{figure}

To reduce the complexity in this search, we introduce a greedy pruning heuristic. At each time step $i$, we limit ourselves to finding only $M_u$ options. As illustrated in Fig.~\ref{fig:pruning}, at each stage we evaluate the incremental cost for all $N_u$ control choices and then prune all but the $M_u$ best. This leads to a computational complexity of only ${O}(N_uM_uT)$. 

\begin{figure*}[htp!]
	\centering
        \begin{subfigure}{0.245\textwidth}
            \centering\includegraphics[width=\textwidth]{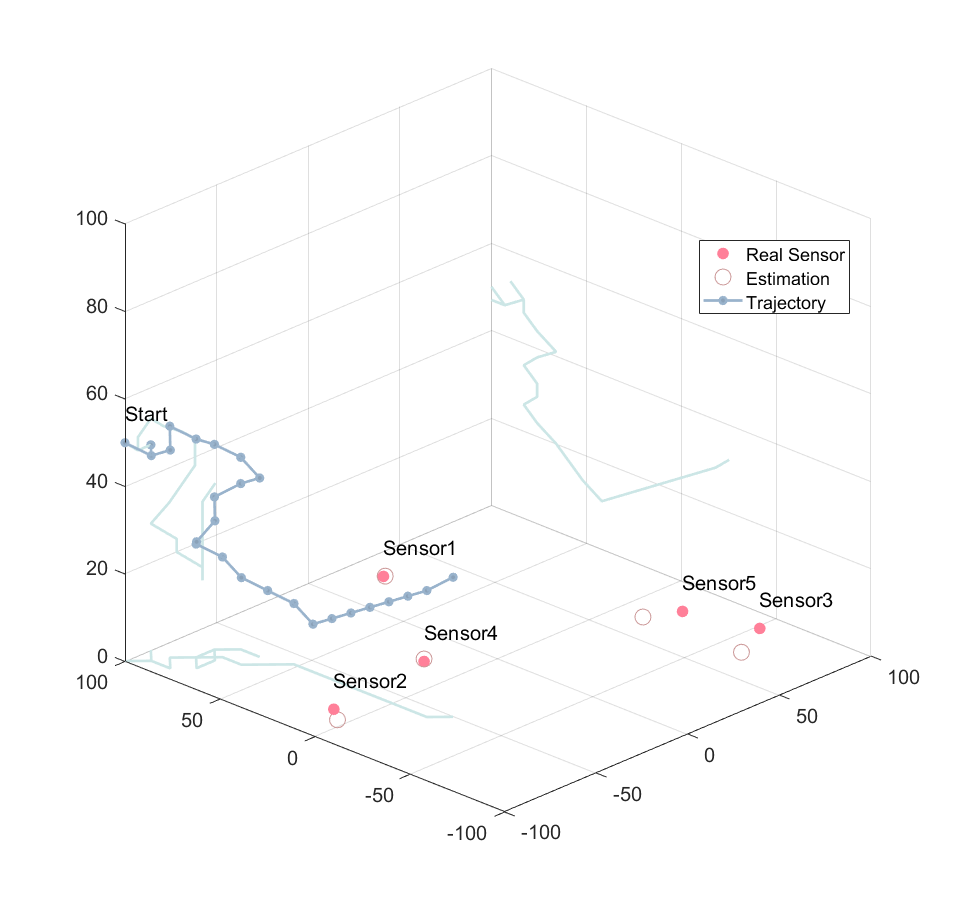}
            \caption{Greedy}
        \end{subfigure}
        \begin{subfigure}{0.245\textwidth}
            \centering\includegraphics[width=\textwidth]{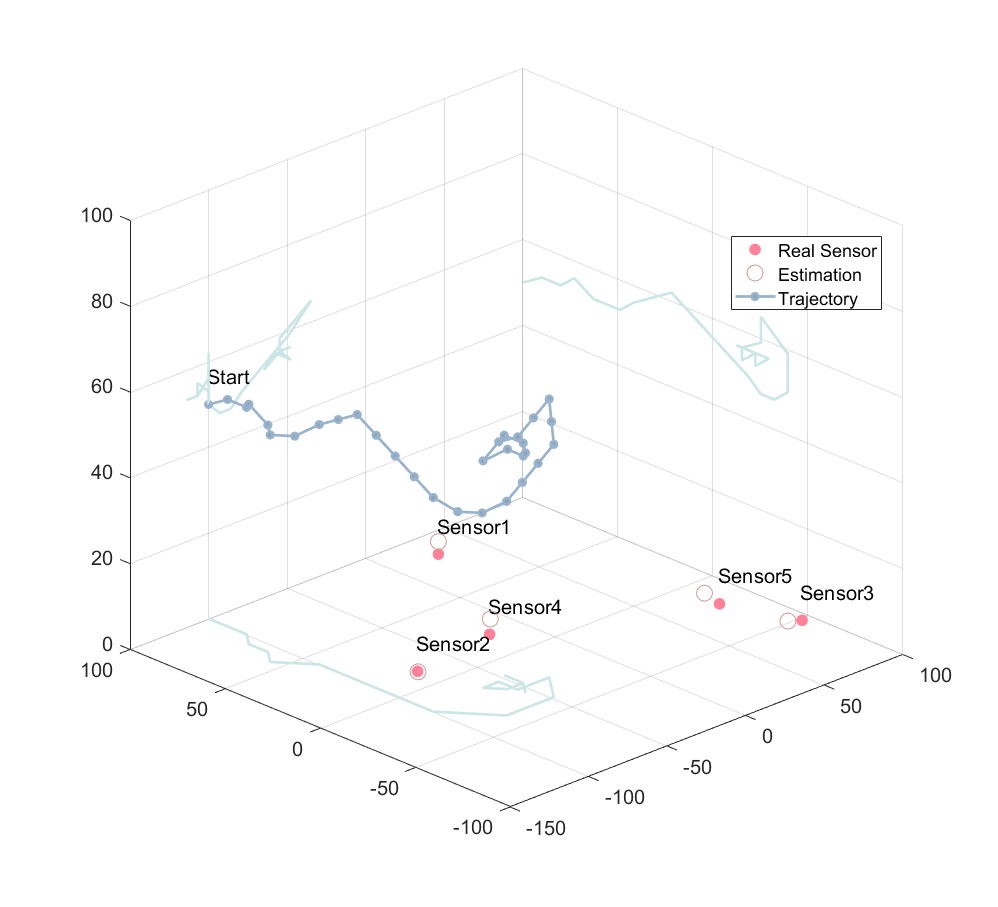}
            \caption{RH with pruning}
        \end{subfigure}	
        \begin{subfigure}{0.245\textwidth}
            \centering\includegraphics[width=\textwidth]{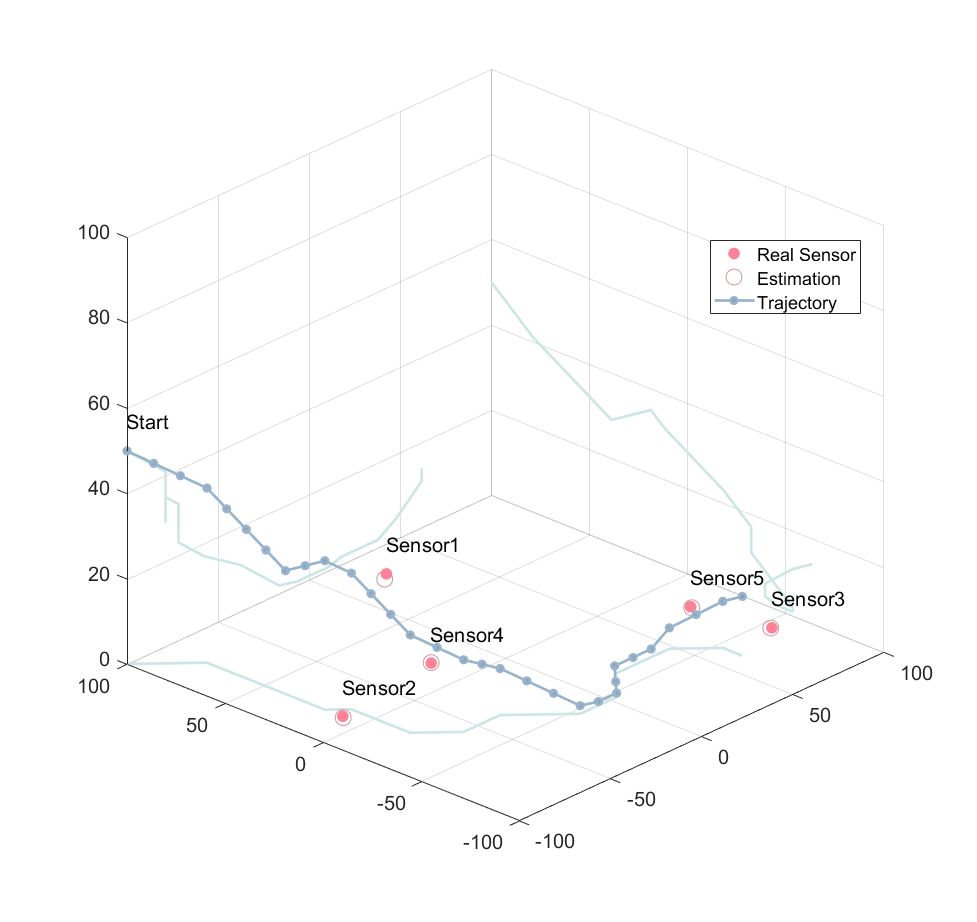}
            \caption{RH DP}
        \end{subfigure}
        \begin{subfigure}{0.245\textwidth}
            \centering\includegraphics[width=\textwidth]{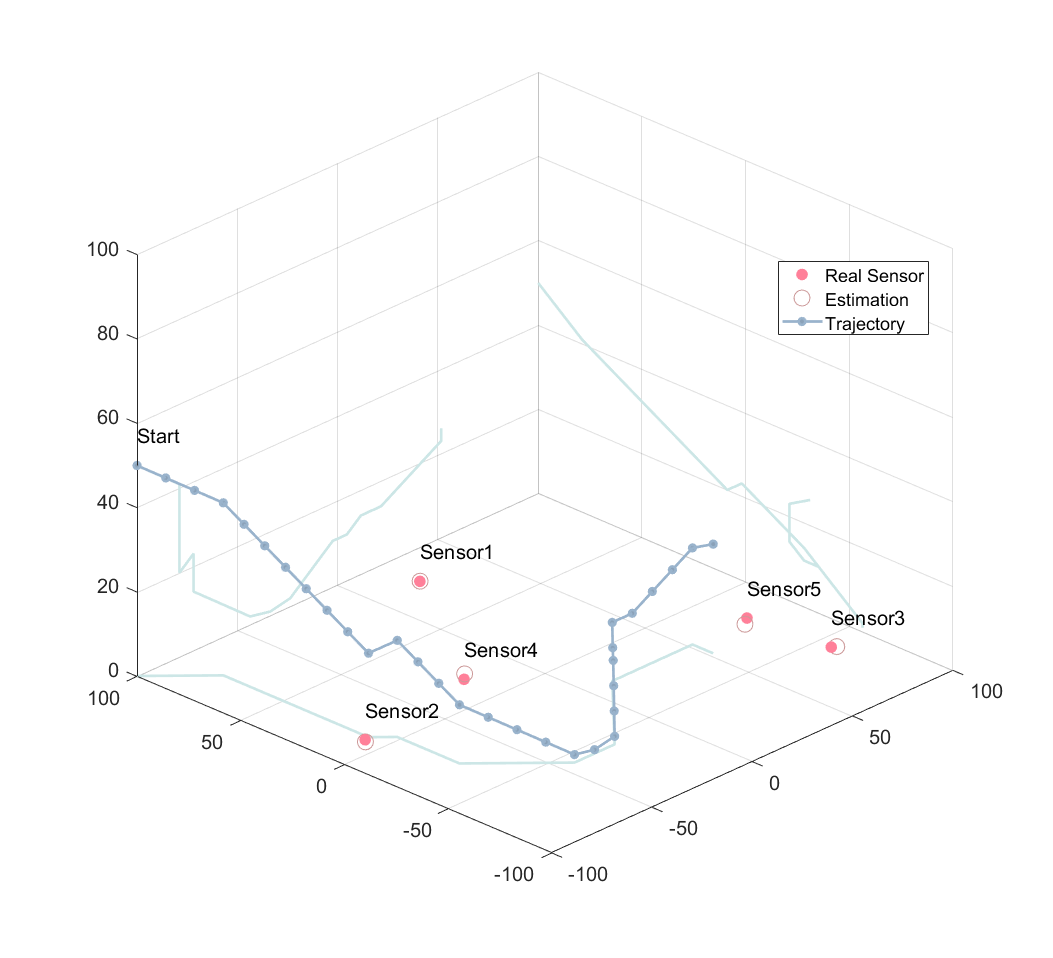}
            \caption{RH DP S}
        \end{subfigure}
        \caption{Simulation run with five sensor nodes under four different approaches with the same initial conditions. The trajectory's 2D projections are also visually presented on the respective axis planes. }
	\label{fig:Motion_Control_Simulation}
\end{figure*}

\subsection{Reward Penalty\label{sec:Penalty}}
As noted above and illustrated in Fig.~\ref{fig:FIM_for_ith_sensor}, using the trace of the inverse of the FIM leads to not care about any individual parameter - achieving high precision about the location of a sensor in one axis at the cost of precision in the other axis is equivalent to achieving equal (but lower) precision in both. In practice, however, it is typically better to have good estimation in all parameters of interest (and to simply exclude parameters you are not interested in ). This can be handled by selecting a different cost function from the FIM (using, for example, minimizing the largest eigenvalue of the inverse), but at the cost of a more complicated function to compute. Here we take a different approach and add a penalty term to the cost function in Problem \ref{prob:RH_MPC_optimization} to encourage minimizing the variance between the diagonal entries. This cost function is
\begin{equation}   
    \label{eq:J2_penalty}
     J'_2 = \rm{tr}(\rm{F}^{-1}_{k+T}) + \beta^{k+T} \cdot Var(Diag(\rm{F}_{k+T}))
\end{equation} 
where $\beta (\beta >1)$ is a user defined coefficient. Notice that as the measurement number $k$ increases, the weight on the variance also increases. This reflects the fact that early on there is little information and thus the estimates, and the resulting FIM expressions, are unreliable and we prioritize simply gaining information. As measurements are accumulated, these estimates become more trustworthy and the algorithm looks to balance over all the parameters.

The complete RH control algorithm for the online signal model estimation is shown in Algorithm 1.

\begin{algorithm}
\caption{Online FIM-based RH Control \& Estimation }
\label{alg:agents_optimization}
\begin{algorithmic}[1]
\State {\bf Input:} $\bm{x}_0,\bm{\hat{\theta}}_0, U, T$ 
\State {\bf Initialization:} $\Phi,\Theta $
\For {$i=0,\dots,T$} :
\State $ \textbf{Measure } \bm{y}_{i} = \Phi(x_i)  \textbf{ from sensor nodes} $
\State $ \textbf{Estimate }  \bm{\hat{\theta}}_{i+1} = \Theta( {x}_0,\dots,{x}_i, {y}_0,\dots,{y}_i, \bm{\hat{\theta}}_{i}) $
\State $ \textbf{Get } u_i \textbf{ by solving Problem 3}$ 
\State $ x_{i+1} = x_i + u_i$
\EndFor
\State {\bf Output:} $ \textbf{Estimation } \bm{\hat{\theta}}_T$
\end{algorithmic}
\end{algorithm}

\section{Simulation and Results}
\label{sec:simulation}

In this section we present simulation results to demonstrate the performance of our algorithm and compare it to two different baselines as well as to different variants of our approach. We consider a scenario with five sensor nodes randomly according to a uniform distribution within a 200 m $\times$ 200 m region with heights ranging from 0 to 10 m to represent hilly terrain. The agent is initialized at one edge of the environment with the starting location of $\begin{bmatrix} 100, -100, 50 \end{bmatrix}$ m. The 24 control actions are defined with a length of $r=10$ m and a vertical move of $h=3$ m. The wireless propagation parameters for each sensor node are chosen from a uniform distribution over the range $\gamma_i \in [5,10]$ and $K_i \in [-30, -10]$ dB. The variance of the noise in the measurements is chosen from a uniform distribution over the range $\sigma_i^2 \in [2,5]$ dB.  All simulations used a total of $N=30$ measurements.

The two baselines considered are a purely random solution where the agent just chooses an action at random at each time step, and a greedy approach where the agent chooses the action that minimizes $J_2$ in the next step. The different variants of the RH approach considered are the DP solution with pruning (RH with pruning), a full DP solution (RH DP), and a DP solution using the modified cost $J_2'$ (RH DP S). In each of these the horizon was set to $T=5$. For RH with pruning we set the number of branches to $M_u=24$.

In Fig. \ref{fig:Motion_Control_Simulation} we show a typical run under each of four controllers (omitting the purely random baseline). The true sensor locations are shown by the solid red circles while the open red circles indicate the estimated locations at the end of 30 time steps. In this particular run, the RH-based algorithms did a better job with localization. It is also interesting to note that the two methods that used full DP performed a more complete exploration of the physical space. 
Of course, the goal is not to explore the physical domain but to accurately estimate the parameters and it is not clear the better exploration necessarily leads to better estimation.

\begin{figure}[htp!]
	\centering
	\includegraphics[width=0.49\textwidth]{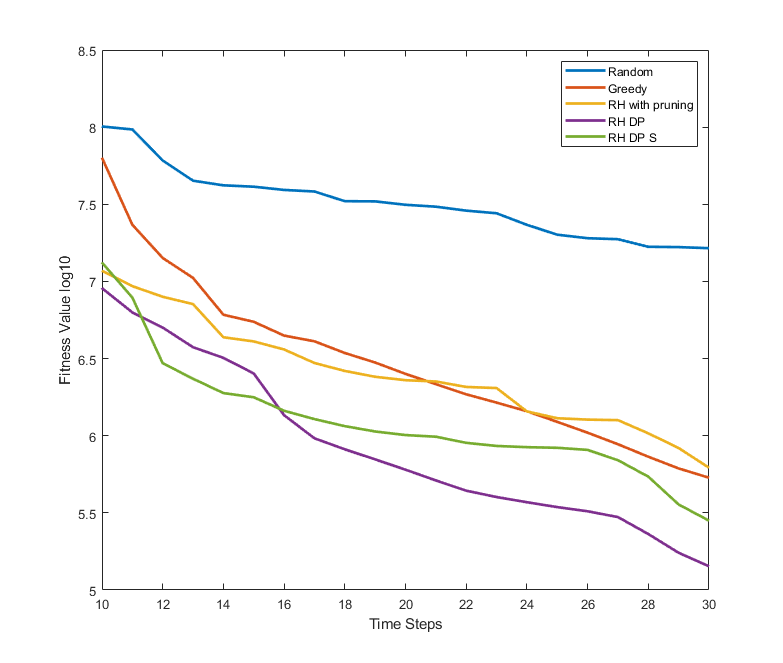}
	\caption{The fitness curves from our algorithm and the other methods in simulation with 5 sensor nodes. The fitness values represent the objective $\rm{tr}(\rm{F}^{-1}_{k}(\bm{\hat{\theta}}))$ in Prob. \ref{prob:basic_online_control_FIM} at $k$th time step.}
	\label{fig:Comparison_Obj_Curve}
\end{figure}

Fig.~\ref{fig:Comparison_Obj_Curve} shows a typical evolution over time of the cost (the trace of the inverse of the FIM) over a run for each of the five algorithms. Note that the figure only shows the evolution after the first 10 measurements. Interestingly, algorithms except for random search seem to show a similar rate of improvement after the first initial steps (though one should keep in mind that this is just from a single run).

\begin{figure*}[htp!]
    \centering
    \begin{subfigure}{0.32\textwidth}
        \centering\includegraphics[width=\textwidth]{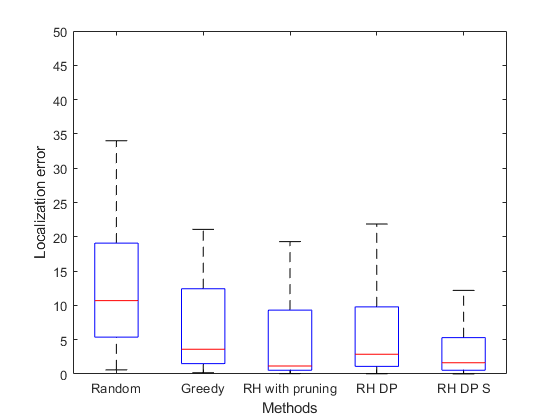}
        \caption{Location Estimation Error}
        \label{fig:Location_error}
    \end{subfigure}
    \begin{subfigure}{0.32\textwidth}
        \centering\includegraphics[width=\textwidth]{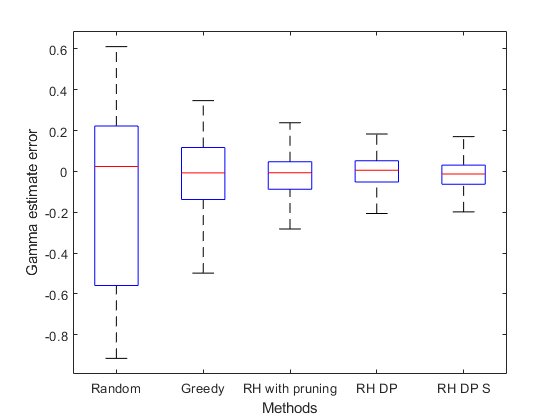}
        \caption{$\gamma$ Estimation Error}
        \label{fig:Gamma_error}
    \end{subfigure}
    \begin{subfigure}{0.31\textwidth}
        \centering\includegraphics[width=\textwidth]{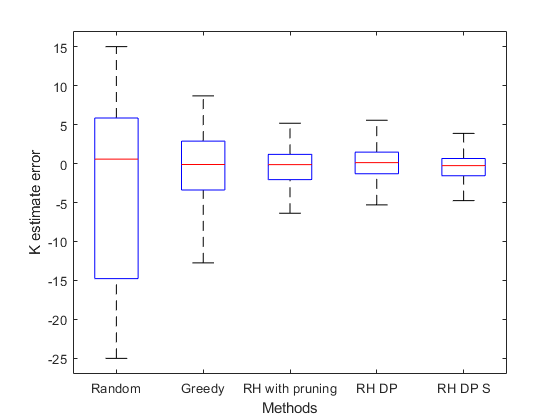}
        \caption{$K$ Estimation Error}
        \label{fig:K_error}
    \end{subfigure}
    \caption{Boxplot of estimation error from different methods for 5 sensors in 50 times simulations. (a) Location error was computed by taking norm of the error in three dimensions. (b) and (c) were values directly taken from the estimation error.}
    \label{fig:Boxplot_estimation_error}
\end{figure*}

To compare performance, we ran each algorithm over 50 realizations. Fig.~\ref{fig:Boxplot_estimation_error} shows the resulting estimation error, with the localization error reported as the three dimensional distance between the estimated and true location. The results indicate that using the RH approach leads to significant improvement in performance over the baseline. The three RH schemes, however, all performed quite similarly in these runs, though it is a topic of ongoing work to explore the space of algorithm hyperparameters for best performance (such as the choice of the value of $\beta$ in $J_2'$ and the horizon $T$).

We also calculated the computation time for one step of the algorithms under all versions, normalizing to the RH DP scheme. The results, shown in Table \ref{table:timings} show that, as expected, the greedy approach is significantly faster than the DP-based solutions, though, surprisingly from Fig.~\ref{fig:Boxplot_estimation_error}, with only limited reduction in performance. Pruning also yielded significant reduction in computation time while also outperforming the greedy approach in terms of estimation error.
\begin{table}[htbp!]
    \caption{Relative one-step computation times}
    \centering
    \scalebox{1.3}{
    \begin{tabular}{l|c}
        \rowcolor{DarkGray} \textcolor{white}{Algorithm} & \textcolor{white}{Time (normalized)} \\ \hline
        Random & 0.223 $\pm$ 0.029 \\ \hline
        Greedy & 0.241 $\pm$ 0.026 \\ \hline
        RH with pruning & 0.651 $\pm$ 0.012 \\ \hline
        RH DP & 1.000 $\pm$ 0.027 \\ \hline
        RH DP S & 2.209 $\pm$ 0.039 \\
        \bottomrule
    \end{tabular}}
    \label{table:timings}
\end{table}

\section{Conclusion and Future Work}
\label{sec:conclusion}

This paper describes an approach to find the best trajectory estimating both the locations of a collection of sensors nodes and the parameters that define the communication model based on measurements of the received signal strength. By optimizing a metric based on the Fisher Information, we seek a trajectory that maximally informs all unknowns. Through simulations, we demonstrate three different variants of the approach and compare them with two simple baselines.

There are several natural extensions to build upon these results. One is to include non line-of-sight (NLoS) measurements into the model. In realistic scenarios, there will always be obstacles in the way, whether they are trees or other natural objects when considering a forest monitoring scenario, or buildings when working with smart cities. Shadowing and multi-path effects can be added to the model, but the lack of prior knowledge of the locations of the sensors makes the problem quite challenging. Another line of work is to jointly optimize the agent path for estimation (explore) as well as for data harvesting (exploit) with a smooth and continuous optimal trajectory which can likely better handle more complicated scenarios.

\section*{Acknowledgements}

This work was funded in part by NSF through grant ECCS-1931600.

\bibliographystyle{IEEEtran}
\balance
\bibliography{references}

\end{document}